\documentclass[a4paper,10pt,twocolumn]{article}
\usepackage{graphicx}
\usepackage{amsmath}
\usepackage{amsbsy}
\usepackage{cite}
\begin{document}

\title{Parameters of Innermost Stable Circular Orbits of Spinning Test Particles: Numerical and Analytical Calculations}

\author{O.Yu. Tsupko$^{1,2}$, G.S. Bisnovatyi-Kogan$^{1,2}$ and P.I. Jefremov$^{3}$
\\[5mm]
\it $^1$Space Research Institute of Russian Academy of Sciences,\\
\it Profsoyuznaya 84/32, Moscow 117997, Russia\\
\it $^2$National Research Nuclear University MEPhI (Moscow Engineering Physics Institute),\\
\it Kashirskoe Shosse 31, Moscow 115409, Russia\\
\it $^3$Center of Applied Space Technology and Microgravity (ZARM),\\
\it University of Bremen, Am Fallturm, 28359 Bremen, Germany\\\\
\it e-mail: tsupko@iki.rssi.ru, gkogan@iki.rssi.ru, paul.jefremow@zarm.uni-bremen.de\\
}

\date{}
\maketitle

\begin{abstract}
The motion of classical spinning test particles in the equatorial plane of a Kerr black hole is considered for the case where the particle spin is perpendicular to the equatorial plane. We review some results of our recent research of the innermost stable circular orbits (ISCO) \cite{Jefremov2015} and present some new calculations. The ISCO radius, total angular momentum, energy, and orbital angular frequency are considered. We calculate the ISCO parameters numerically for different values of the Kerr parameter $a$ and investigate their dependence on both black hole and test particle spins. Then we describe in details how to calculate analytically small-spin corrections to the ISCO parameters for an arbitrary values of $a$. The cases of Schwarzschild, slowly rotating Kerr and extreme Kerr black hole are considered. The use of the orbital angular momentum is discussed. We also consider the ISCO binding energy. It is shown that the efficiency of accretion onto an extreme Kerr black hole can be larger than the maximum known efficiency (42 $\%$) if the test body has a spin.
\end{abstract}

\section{Introduction}

\subsection{Innermost stable circular orbits in General Relativity}

\begin{figure}
\centerline{\hbox{\includegraphics[width=0.45\textwidth]{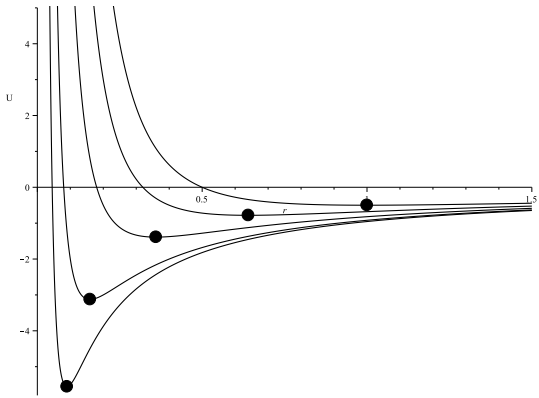}}} \caption{Newtonian effective potential $U =  - M/r +  L^2/2r^2$ for a test particle moving in the gravitational field of a central body with mass $M$, for different values of $L$ ($L$ is the angular momentum per unit mass, $G=1$). Solid circles show positions of minima corresponding to stable circular orbits.} \label{fig-U-Newt}
\end{figure}

Let us consider a classical test non-spinning particle moving at stable circular orbit around a central massive body. In Newtonian theory, this orbit can have arbitrary radius. This follows from the fact that the effective potential of particle always has minimum, for any value of particle angular momentum, and if angular momentum tends to zero the radius of stable circular orbit goes to zero too \cite{Zeld-Novikov1971}, see Fig. \ref{fig-U-Newt}. All circular orbits are stable till zero radius, and in Newtonian theory there is no minimum radius of stable circular orbit \cite{Kaplan}.

\begin{figure}
\centerline{\hbox{\includegraphics[width=0.45\textwidth]{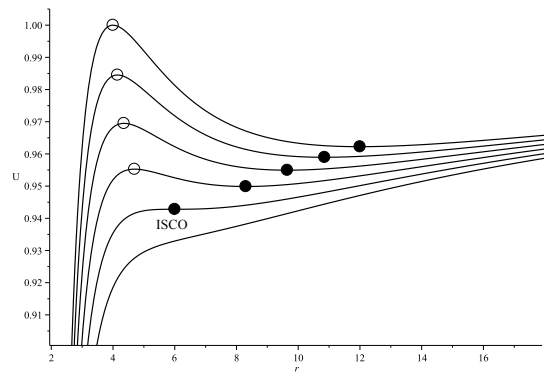}}} \caption{Effective potential (per unit particle rest mass) for motion in Schwarzschild metric $U_{Schw}= \sqrt{\left( 1 -\frac{2M}{r} \right) \left( 1 +\frac{L^2}{r^2} \right)}$ for different values of $L$ ($L$ is the angular momentum per unit particle rest mass, $G=1$). Maxima of potential are shown by circles, and minima are shown by solid circles.} \label{fig-U-Schw}
\end{figure}

In General Relativity (GR) the situation is different. The effective potential has a more complicated form, depending on the particle angular momentum, see Fig. \ref{fig-U-Schw} for the potential in the Schwarzschild metric. For large values of the angular momentum the effective potential has two extrema: maximum which corresponds to unstable circular orbit, and minimum which corresponds to stable circular orbit. With decreasing of angular momentum, radii of unstable and stable circular orbits become closer to each other. When angular momentum reach a boundary value, two extrema of effective potential merge into one inflection point. This point corresponds to minimal possible radius of stable circular orbit. Such orbit is called the innermost stable circular orbit (ISCO). Further decreasing of angular momentum leads to potential without extrema. For these values of angular momentum neither type of finite motion is possible.

\begin{figure}
\centerline{\hbox{\includegraphics[width=0.45\textwidth]{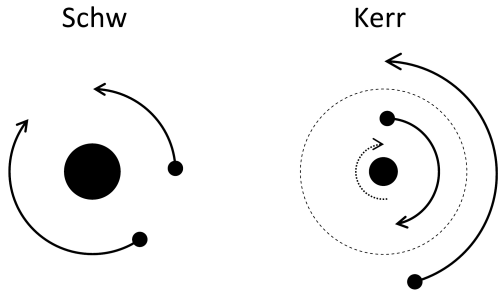}}} \caption{Innermost stable circular orbits in Schwarzschild and Kerr metric. In Kerr metric radius of ISCO depends on direction of orbital motion of test particle.} \label{fig-schw-kerr}
\end{figure}

Radius and other values of the ISCO parameters (total angular momentum, energy, orbital angular frequency) are different in different metrics. For the Schwarzschild background the radius of ISCO equals to $6M$\footnote{In this paper we use the system of units where $G=c=1$, the Schwarzschild radius $R_S=2M$, and other physical quantities which will be introduced further have the following dimensionalities: $[L]=[M]$, $[J]=[M]$, $[E]=1$, $[a]=[M]$, $[s]=[M]$.} , it was found by Kaplan \cite{Kaplan}, see also \cite{LL2}. In the Kerr space-time circular motion is possible only in the equatorial plane of BH and the radius of ISCO depends on the direction of motion of the particle in comparison with the direction of BH rotation, whether they co-rotate or counter-rotate. Co-rotation and counter-rotation cases correspond to parallel and antiparallel orientation of vectors of the orbital angular momentum of the particle and the BH angular momentum. In a case of orbital co-rotation the ISCO radius becomes smaller than $6M$, in case of counter-rotation -- bigger, see Fig. \ref{fig-schw-kerr}. For the case of the extreme Kerr background the difference between these two variants is quite considerable: we have $9M$ for the antiparallel and $M$ for the parallel orientation. The parameters of ISCO in the Kerr space-time for a non-spinning particle were obtained in works of Ruffini \& Wheeler \cite{Ruffini1971} and Bardeen, Press \& Teukolsky \cite{Bardeen1972}. This problem is described at length, for example, in the textbook by Hobson \textit{et al.} \cite{Hobson}.

\subsection{Spinning particles in General Relativity}

In GR a rotation of the central gravitating body influences a motion of the particle orbiting it. Due to this reason the orbits of test particles in Kerr metric differ from orbits in Schwarzschild metric. When, in turn, a test particle has spin as well, it will also influence the particle's orbit. In particular, the motion of a spinning particle will differ from the non-spinning one even in the Schwarzschild background.

The problem of the motion of a classical spinning test body in GR was considered in papers of Mathisson \cite{Mathisson1937}, Papapetrou \cite{Papapetrou1951a} and Dixon \cite{Dixon1970a, Dixon1970b, Dixon1978}, using different techniques. The equations of motion of a spinning test particle in a given gravitational field were derived in different forms; they are now referred to as Mathisson-Papapetrou-Dixon equations. From these equations it follows that the motion of the centre of mass and the particle rotation are connected to each other, and when the particle has spin the orbits will differ from geodesics of a spinless massive particle.

\subsection{The ISCO of spinning particles}

Influence of a spin on the orbits in the Schwarzschild metric was investigated in the paper of Corinaldesi and Papapetrou \cite{Papapetrou1951b}, and in the paper of Micoulaut \cite{Micoulaut1967}. A motion of a spinning test particle in Kerr metric was considered by Rasband \cite{Rasband1973} and Tod, de Felice \& Calvani \cite{Tod1976}, in particular ISCO radius was calculated numerically, see also papers of Abramowicz \& Calvani \cite{Abramowicz1979} and Calvani \cite{Calvani1980}. Subsequently the number of works on this subject were published \cite{Hojman1977,Suzuki1997, Suzuki1998, SaijoMaeda1998, TanakaMino1996, Apostolatos1996, Semerak1999, Semerak2007, Plyatsko2012a, Plyatsko2012b, Plyatsko2013, Bini2004a, Bini2004b, Bini2011a, Bini2011b, BiniDamour2014, Damour2008, Faye2006, Favata2011, Steinhoff2011, Steinhoff2012, Hackmann2014,Kunst2015,Putten1999}.

The detailed derivation of the equations of motion of spinning particle in Kerr space-time is presented in the work of Saijo \textit{et al.} \cite{SaijoMaeda1998}. Method of calculation of ISCO parameters of spinning particle moving in Kerr metric is presented in details \cite{Favata2011}. Linear corrections in spin for the ISCO parameters in Schwarzschild metric have been found by Favata \cite{Favata2011}.

In paper of Jefremov, Tsupko \& Bisnovatyi-Kogan \cite{Jefremov2015} we have analytically obtained the small spin corrections for the ISCO parameters for the Kerr metric at arbitrary value of Kerr parameter $a$. The cases of Schwarzschild, slowly rotating and extreme Kerr black hole were considered in details. For a slowly rotating black hole the ISCO parameters are obtained up to quadratic in $a$ and particle's spin $s$ terms. For the extreme $a=M$ and almost extreme $a=(1-\delta)M$ Kerr BH we succeeded to find the exact analytical solution for the ISCO parameters for arbitrary spin, with only restrictions connected with applicability of Mathisson-Papapetrou-Dixon equations. It has been shown that the limiting values of ISCO radius and frequency for $a=M$ do not depend on the particle's spin while values of energy and total angular momentum do depend on it.

In this work we review some results of our recent research of innermost stable circular orbits (ISCO) \cite{Jefremov2015} and present some new calculations. ISCO radius, total angular momentum, energy, orbital angular frequency are considered. We calculate the ISCO parameters numerically for different values of Kerr parameter $a$ and investigate their dependence on both black hole and test particle spins. Then we describe in details how to calculate analytically small-spin corrections to ISCO parameters for arbitrary values of $a$, presenting our formulae in different forms.

\section{The motion of a spinning test body in the equatorial plane of a Kerr black hole} \label{section-MPD}

In the present treatment of the problem of spinning body motion in GR we use the so-called "pole-dipole" approximation \cite{Papapetrou1951a}, in the frame of which the motion is described by the Mathisson-Papapetrou-Dixon (MPD) equations \cite{Papapetrou1951a,Mathisson1937,Dixon1970a,Dixon1970b, Dixon1978, SaijoMaeda1998}:
\begin{equation}
\begin{split}
&\frac{Dp^\mu}{D\tau}=-\frac{1}{2}R^{\mu}{}_{\nu \rho \sigma}v^{\nu}S^{\rho \sigma} ,\\
&\frac{DS^{\mu \nu}}{D\tau}=p^\mu v^\nu - p^\nu v^\mu.
\label{MPD}
\end{split}
\end{equation}
Here $D/D \tau$ is a covariant derivative along the particle trajectory, $\tau$ is an affine parameter of the orbit \cite{SaijoMaeda1998}, $R^{\mu}{}_{\nu \rho \sigma}$ is the Riemannian tensor, $p^\mu$ and $v^\mu$ are 4-momentum and 4-velocity of a test body, $S^{\rho \sigma}$ is its spin-tensor. The equations were derived under the assumption that characteristic radius of the spinning particle is much smaller than the curvature scale of a background spacetime \cite{SaijoMaeda1998} (see also \cite{Rasband1973}, \cite{Apostolatos1996}) and the mass of a spinning body is much less than that of BH.

It is known, however, that these equations are incomplete, because they do not define which point on the test body is used for spin and trajectory measurements. Therefore we need some extra condition (`spin supplementary condition') to do that and to close the system of equations \cite{Papapetrou1951b}. We use the condition of Tulczyjew \cite{Tulczyjew1959} given by
\begin{equation}
p_\mu S^{\mu \nu}=0.
\label{SSC}
\end{equation}

The system of equations (\ref{MPD}) with (\ref{SSC}) in a general space-time admits two conserved quantities: particle's mass $m^2= -p^{\mu}p_{\mu}$ and the magnitude of its specific spin $s^2= S^{\mu \nu}S_{\mu \nu}/(2m^2)$, see \cite{SaijoMaeda1998}.

We will consider the motion of a spinning particle in the equatorial plane of Kerr metric ($\theta=\pi/2$),
which is given in Boyer-Lindquist coordinates by \cite{LL2, Hobson}
\begin{equation}
\begin{split}
ds^2 & = - \left( 1 -\frac{2M r}{\Sigma} \right) dt^2 - \\
&- \frac{4M a r \sin^2 \theta}{\Sigma} dt \ d\varphi + \frac{\Sigma}{\Delta}dr^2 + \Sigma \ d\theta^2 + \\
&+ \left( r^2 +a^2  +\frac{2M ra^2 \sin^2\theta}{\Sigma} \right)\sin^2\theta \ d\varphi^2,
\end{split}
\label{}
\end{equation}
where $a$ is the specific angular momentum of a black hole, $\Sigma \equiv r^2 + a^2 \cos^2\theta$, $\Delta \equiv r^2 - 2 Mr + a^2$. In this case there are two additional conserved quantities: total energy of the particle and the projection of its total angular momentum onto $z$-axis.

In case of the motion of a spinning particle in the equatorial plane, the angular momentum of a spinning particle is always perpendicular to the equatorial plane \cite{SaijoMaeda1998}. Therefore we can describe the test particle spin by only one constant $s$ which is the specific spin angular momentum of the particle. Value $|s|$ indicates the magnitude of the spin and $s$ itself is its projection on the $z$-axis. It is more obvious to think of the spin in terms of the particle's spin angular momentum $\mathbf{S_1}=sm\mathbf{\hat{z}}$ which is parallel to the BH spin angular momentum $\mathbf{S_2}=aM\mathbf{\hat{z}}$, when $s>0$, and antiparallel, when $s<0$. Here $\mathbf{\hat{z}}$ is a unit vector in the direction of the $z$-axis and $m$ is a mass of the particle \cite{SaijoMaeda1998}, \cite{Favata2011}.

Saijo \textit{et al} \cite{SaijoMaeda1998} have derived the equations of motion of a spinning test particle for the equatorial plane of Kerr BH. The equations of motion for the variables $r$, $t$, $\varphi$ in this case have the form \cite{SaijoMaeda1998}
\begin{equation}
\begin{split}
& (\Sigma_s \Lambda_s \dot r)^2 =R_s,\\
& \Sigma_s \Lambda_s \dot t =a\left( 1 +\frac{3Ms^2}{r \Sigma_s}\right)\left[ J -(a+s)E \right] +\\
& +\frac{r^2 +a^2}{\Delta}P_s,\\
& \Sigma_s \Lambda_s \dot \varphi =\left( 1 +\frac{3Ms^2}{r \Sigma_s}\right)\left[ J -(a+s)E \right] +\frac{a}{\Delta}P_s.\\
\end{split}
\label{spin-eqs}
\end{equation}
where
\begin{equation}
\begin{split}
& \Sigma_s= r^2 \left(1 -\frac{Ms^2}{r^3}\right),\\
& \Lambda_s= 1 - \frac{3 M s^2 r [-(a + s) E +J]^2}{\Sigma_s^3},\\
&R_s = P_s^2 - \Delta \left\{ \frac{\Sigma_s^2}{r^2} + [-(a + s) E +J]^2 \right\}, \\
&P_s = \left[r^2 + a^2 + \frac{a s (r + M)}{r} \right]  E  -   \left(a + \frac{M s}{r} \right) J.
\end{split}
\end{equation}
Here $\dot{x} \equiv dx/d \tau$ and the affine parameter $\tau$ is normalised as $p^{\nu} v_{\nu} =-m$ \cite{SaijoMaeda1998}; $E$ is the conserved energy per unit particle rest mass, and $J=J_z$ is the conserved total angular momentum per unit particle rest mass which is collinear to the spin of a BH.

We can write the equation for radial motion in the form \cite{Favata2011}, \cite{Rasband1973}
\begin{equation}
(\Sigma_s \Lambda_s \dot r)^2 =\alpha_s E^2 -2\beta_s E +\gamma_s,
\label{rad-motion}
\end{equation}
where
\begin{equation}
\begin{split}
&\alpha_s = \left[r^2 + a^2 + \frac{a s (r + M)}{r} \right]^2 - \Delta (a + s)^2,\\
&\beta_s = \left[ \left(a + \frac{M s}{r} \right) \left(r^2 + a^2 + \frac{a s (r + M)}{r} \right) - \right. \\
&\left. - \Delta (a + s) \right]J,\\
&\gamma_s = \left(a + \frac{M s}{r} \right)^2 J^2 - \Delta \left[r^2 \left(1 - \frac{M s^2}{r^3}\right)^2 + J^2 \right].
\end{split} \label{abc-spin}
\end{equation}
We can consider the whole right-hand side of (\ref{rad-motion}) as an effective potential. For the matter of convenience, let us further divide it by $r^4$ and define the effective potential as
\begin{equation}
V_s(r;J,E)= \frac{1}{r^4} (\alpha_s E^2 -2\beta_s E +\gamma_s).
\label{V-spin}
\end{equation}

\section{Numerical calculation of ISCO parameters}

The equations which define circular orbits are given by a system
\begin{equation}
\left\{
\begin{aligned}
V_s &= 0 \, ,\\
\frac{dV_s}{dr} &=0 \, .\\
\end{aligned}
\right.
\label{eff12}
\end{equation}
In order to find the last stable orbit (ISCO) we need to demand additionally that the second derivative of the effective potential vanishes:
\begin{equation}
\begin{aligned}
&\frac{d^2 V_s}{dr^2} = 0.\\
&\\
\end{aligned}
\label{eff3}
\end{equation}

For the sake of convenience, we change variables and work not with $r$ and $J$ but with $u=1/r$ and $x=J-aE$, so the function $V_s(u; x,E)$ will be used. In new variables (see \cite{Jefremov2015} for details), system of equations determining ISCO have the form
\begin{equation}
\left\{
\begin{aligned}
V_s &= 0 \, ,\\
\frac{dV_s}{du} &=0 \, ,\\
\frac{d^2 V_s}{du^2} &= 0 \, .\\
\end{aligned}
\right.
\label{n}
\end{equation}
The explicit form of these equations is \cite{Jefremov2015}:
%\begin{widetext}
\begin{equation}
\begin{aligned}
&(1 + 2 a s u^2 - s^2 u^2 + 2 M s^2 u^3) E^2 + \\
&+(-2 a u^2 x + 2 s u^2 x - 6 M s u^3 x - 2 a M s^2 u^5 x) E -\\ &-1 + 2 M u - a^2 u^2 + 2 M s^2 u^3 -\\
&- 4 M^2 s^2 u^4 + 2 a^2 M s^2 u^5 - M^2 s^4 u^6 +\\
&+ 2 M^3 s^4 u^7 - a^2 M^2 s^4 u^8 - u^2 x^2 + 2 M u^3 x^2 +\\
&+ 2 a M s u^5 x^2 + M^2 s^2 u^6 x^2= 0 \, ;\\
&(4 a s u - 2 s^2 u + 6 M s^2 u^2) E^2 + \\
&+ (-4 a u x + 4 s u x - 18 M s u^2 x - 10 a M s^2 u^4 x) E +\\
&+2 M - 2 a^2 u + 6 M s^2 u^2 - 16 M^2 s^2 u^3 +\\
& + 10 a^2 M s^2 u^4 - 6 M^2 s^4 u^5 + 14 M^3 s^4 u^6 -\\
&- 8 a^2 M^2 s^4 u^7 - 2 u x^2 + 6 M u^2 x^2 +\\
&+ 10 a M s u^4 x^2 + 6 M^2 s^2 u^5 x^2=0 \, ;\\
&(4 a s - 2 s^2 + 12 M s^2 u) E^2 + \\
&+ (-4 a x + 4 s x - 36 M s u x - 40 a M s^2 u^3 x)E -\\ 
&- 2 (a^2 - 6 M s^2 u + 24 M^2 s^2 u^2 - 20 a^2 M s^2 u^3 +\\
&+ 15 M^2 s^4 u^4 - 42 M^3 s^4 u^5 +28 a^2 M^2 s^4 u^6 + x^2- \\ 
& - 6 M u x^2 - 20 a M s u^3 x^2 - 15 M^2 s^2 u^4 x^2) =0 \, .
\end{aligned}
\label{system}
\end{equation}
%\end{widetext}

These three equations form a closed system for three parameters of ISCO $E$, $x$ and $u$, which are dependent only on the Kerr parameter $a$ and particle's spin $s$. This system can be used for numerical calculation of $E$, $x$, $u$ of ISCO at given $a$ and $s$. Then, values of $r$ and $J$ can be found numerically by using $r=1/u$ and $J=x+aE$.

Another important characteristics of the particle circular motion is its angular velocity. The orbital angular frequency of the particle at the ISCO, as seen from an observer at infinity, is defined as
\begin{equation}
\Omega \equiv \frac{d \varphi / d\tau}{ dt / d\tau}.
\label{Omega-definition}
\end{equation}
The values $d \varphi / d\tau$ and $dt / d\tau$ are found from the second and the third equations in (\ref{spin-eqs}), where we should substitute values of $r$, $E$ and $J$ at a given orbit. To find the ISCO frequency $\Omega_{\mathrm{\, ISCO}}$ we need to use the ISCO values of $r$, $E$ and $J$, see \cite{Favata2011}.

At given $a$ and $s$ the system lead to solutions both for co-rotating and counter-rotating case. Corotation means parallel orientation of particle's angular momentum $\mathbf{J}$ and BH spin $\mathbf{a}$, $J>0$; counter-rotation means antiparallel orientation, $J<0$. The $z$-axis is chosen to be parallel to BH spin $\mathbf{a}$, so $a$ is positive or equal to zero. Spin $s$ is the projection of spin on the $z$-axis and can be positive (spins of particle and BH are parallel) or negative (antiparallel).

\begin{figure}
\centerline{\hbox{\includegraphics[width=0.45\textwidth]{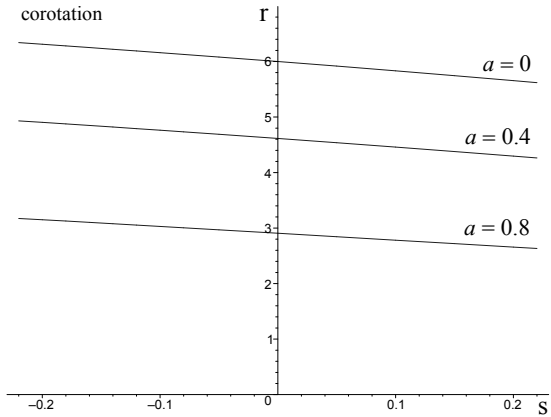}}} \caption{Radius of ISCO for the case of corotation: angular momentum of black hole and total angular momentum of particle are parallel, $J>0$. All values are in units of $M$. See also Figure 3 in paper of Suzuki and Maeda \cite{Suzuki1998}.} \label{fig-r-co}
\end{figure}

\begin{figure}
\centerline{\hbox{\includegraphics[width=0.45\textwidth]{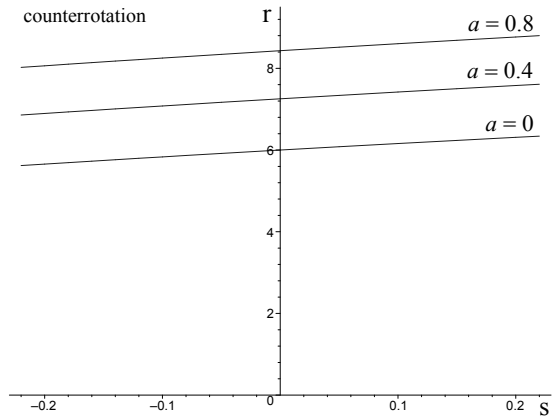}}} \caption{Radius of ISCO for the case of counterrotation: angular momentum of black hole and total angular momentum of particle are antiparallel, $J<0$. All values are in units of $M$.} \label{fig-r-counter}
\end{figure}

Specifying $a$ and $s$, we can find $r_{\mathrm{\, ISCO}}$, $E_{\mathrm{\, ISCO}}$ and $J_{\mathrm{\, ISCO}}$ numerically by solving system (\ref{system}), all parameters are in units of $M$. Results for the radius calculation are presented in Figures \ref{fig-r-co} and \ref{fig-r-counter}. For $a=0$ and non-spinning particle ($s=0$) radius equals to $6M$. Increase of $a$ leads to decrease of $r_{\mathrm{\, ISCO}}$ for co-rotating case and to increase of $r_{\mathrm{\, ISCO}}$ for counter-rotating case.

\begin{figure}
\centerline{\hbox{\includegraphics[width=0.45\textwidth]{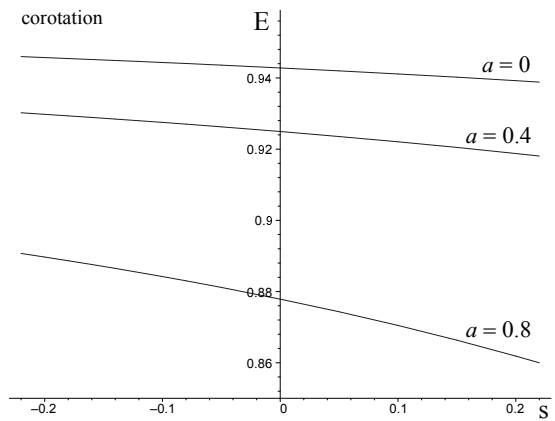}}} \caption{Energy of ISCO for the case of corotation.} \label{fig-E-co}
\end{figure}

\begin{figure}
\centerline{\hbox{\includegraphics[width=0.45\textwidth]{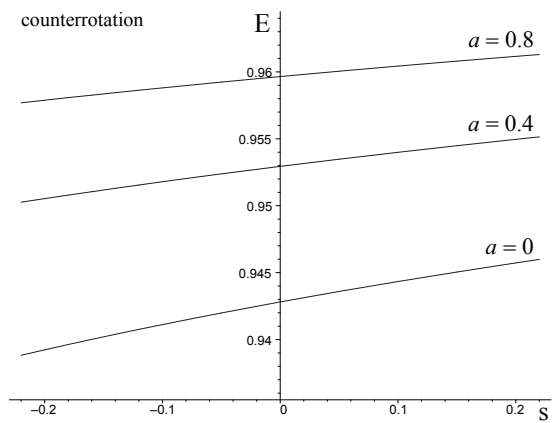}}} \caption{Energy of ISCO for the case of counterrotation.} \label{fig-E-counter}
\end{figure}

Calculations of ISCO energy are shown in Figures \ref{fig-E-co} and \ref{fig-E-counter}. For $a=0$ and $s=0$ the energy equals to $2\sqrt{2}/3$.

\begin{figure}
\centerline{\hbox{\includegraphics[width=0.45\textwidth]{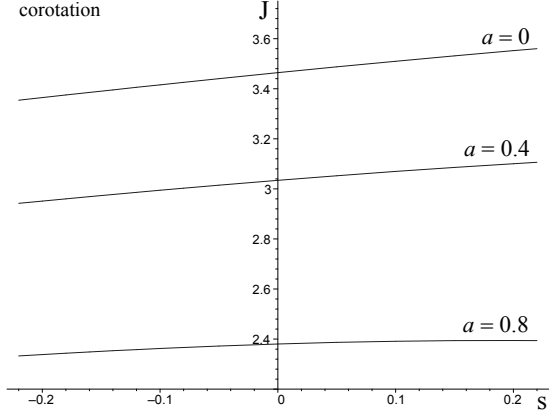}}} \caption{Total angular momentum of ISCO for the case of corotation.} \label{fig-J-co}
\end{figure}

\begin{figure}
\centerline{\hbox{\includegraphics[width=0.45\textwidth]{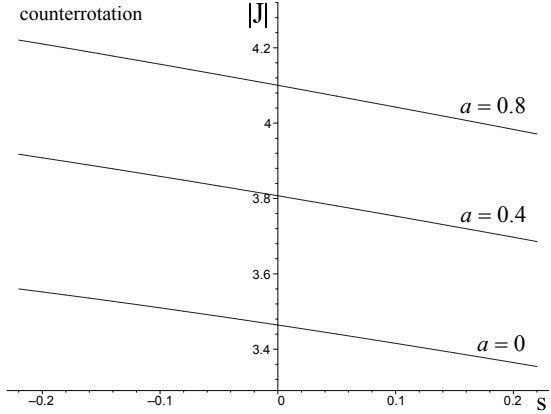}}} \caption{Total angular momentum (absolute value) of ISCO for the case of counterrotation.} \label{fig-J-counter}
\end{figure}

Calculations of ISCO total angular momentum are presented in Figures \ref{fig-J-co} and \ref{fig-J-counter}. For $a=0$ and $s=0$ the magnitude of angular momentum equals to $2\sqrt{3}M$.

\begin{figure}
\centerline{\hbox{\includegraphics[width=0.45\textwidth]{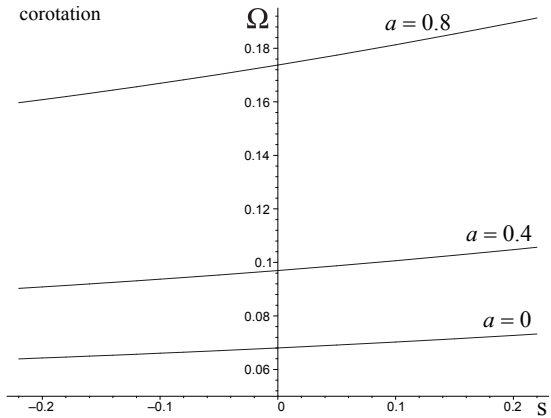}}} \caption{Orbital angular frequency of ISCO for the case of corotation.} \label{fig-Omega-co}
\end{figure}

\begin{figure}
\centerline{\hbox{\includegraphics[width=0.45\textwidth]{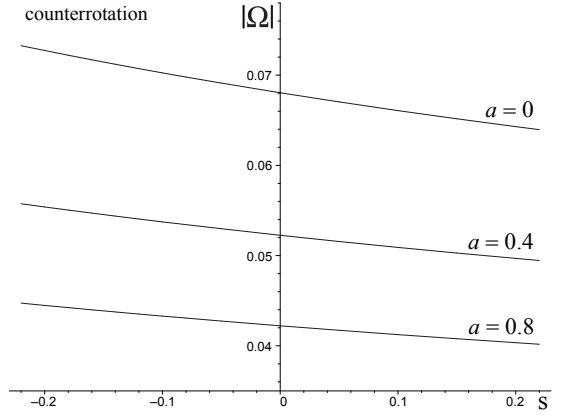}}} \caption{Orbital angular frequency (absolute value) of ISCO for the case of counterrotation.} \label{fig-Omega-counter}
\end{figure}

Calculations of ISCO angular frequency are shown in Figures \ref{fig-Omega-co} and \ref{fig-Omega-counter}. For $a=0$ and $s=0$ the magnitude of angular frequency equals to $1/6\sqrt{6}M$.

\section{Analytical calculation of ISCO parameters, small-spin corrections}

In paper \cite{Jefremov2015} we have derived linear small-spin corrections for ISCO parameters for arbitrary $a$. Parameters are written there with using variable $u_0=1/r_0$ which is inverse radius of ISCO for non-spinning particle. The scheme of calculation of all parameters are presented there, see text after formula (61) in \cite{Jefremov2015}. Here, we rewrite all formulas in terms of $r_0$.

The scheme of calculation of ISCO parameters with linear small-spin corrections:

(i) Radius of ISCO. We need to solve equation for ISCO radius $r_0$ of non-spinning particle
\begin{equation}
r_0^2 - 6Mr_0 - 3a^2 \mp 8a \sqrt{Mr_0} =0.
\label{eq-r0}
\end{equation}
and find $r_0$. In this equation and all formulas below the upper sign corresponds to the antiparallel orientation of particle's angular momentum $\mathbf{J}$ and BH spin $\mathbf{a}$ (counter-rotation, $J<0$), the lower -- to the parallel one (corotation, $J>0$). Solution of (\ref{eq-r0}) can be found analytically, see \cite{Bardeen1972}. To avoid large formulas, we write all other unknowns not as explicit functions of $a$ but as explicit functions of $a$ and $r_0$, keeping in mind that $r_0$ can be found from Eq. (\ref{eq-r0}) at arbitrary $a$. We need to notice that representation of all unknowns via $r_0$ given below could be rewritten in a different form using (\ref{eq-r0}), see \cite{Jefremov2015} for different representations.

Linear correction is
\begin{equation}
r_1 = \frac{4}{r_0} (a \pm \sqrt{Mr_0}) .
\end{equation}
Finally radius of ISCO for given $a$ is
\begin{equation}
r_{\mathrm{\, ISCO}} = r_0 + s r_1 \, .
\end{equation}

(ii) Energy of ISCO is:
\begin{equation}
E_{\mathrm{\, ISCO}} = E_0 + s E_1 ,
\end{equation}
\begin{equation}
E_0 = \sqrt{1-\frac{2M}{3r_0}} ,
\end{equation}
\begin{equation}
E_1 = \pm \frac{1}{\sqrt{3}} \frac{M}{r_0^2} .
\end{equation}

(iii) Total angular momentum of ISCO is:
\begin{equation}
J_{\mathrm{\, ISCO}} = J_0 + s J_1 ,
\end{equation}
\begin{equation}
J_0 = \mp \frac{r_0}{\sqrt{3}} + a \sqrt{1-\frac{2M}{3r_0}} ,
\end{equation}
\begin{equation}
J_1 = \frac{2\sqrt{M} r_0^{3/2} \pm  a(3r_0 +M) }{\sqrt{3} r_0^2} .
\end{equation}

(iv) Orbital angular frequency of ISCO is:
\begin{equation}
\Omega_{\mathrm{\, ISCO}} = \Omega_0 + s \Omega_1 ,
\end{equation}
\begin{equation}
\Omega_0 = \frac{\sqrt{M}}{a\sqrt{M} \mp  r_0^{3/2}   } ,  
\end{equation}
\begin{equation}
\Omega_1 = \frac{9 \sqrt{M} ( \sqrt{r_0 M} \pm a ) }{2 \sqrt{r_0} \left(r_0^{3/2} \mp a \sqrt{M} \right)^2 } .
\end{equation}

In work of Favata \cite{Favata2011} the shift in the ISCO due to the spin of the test-particle is calculated numerically, see right picture on Fig. 2 in \cite{Favata2011} and the sixth column in Table I in \cite{Favata2011}. Our analytical results (with appropriate change of variables) agree with calculations of paper of Favata: value of $\Omega_0$ for given $a$ (see first column in Table I; note that BH spin can be negative in this Table, it corresponds to case of counterrotation in our work) give numbers in second column of Table I, value of $\Omega_1/\Omega_0$ give numbers in six column.

Now let us consider particular cases.

Results for the Schwarzschild case are \cite{Jefremov2015}:
\begin{equation}
\begin{aligned}
J_{\mathrm{\, ISCO}}&=2 \sqrt{3} M + \frac{\sqrt{2}}{3}s_J ,\\
E_{\mathrm{\, ISCO}}&= \frac{2 \sqrt{2}}{3} -\frac{1}{36 \sqrt{3}}\frac{s_J}{M} ,\\
r_{\mathrm{\, ISCO}}&= 6M -2\sqrt{\frac{2}{3}}s_J ,\\
\Omega_{\mathrm{\, ISCO}}&= \frac{1}{6 \sqrt{6}M} +\frac{s_J}{48 M^2} .\\
\end{aligned}
\label{Schw}
\end{equation}
Here instead of $s$, which is the projection on $z$-axis that does not unabiguously correspond to any physical direction in Schwarzschild case, we use $s_J$ which is the projection of particle's spin upon the direction of $\mathbf{J}$ and is positive when the particle's spin is parallel to it and negative when it is antiparallel. Value $J$ is considered as positive in this case. Small-spin corrections for Schwarzschild metric were derived by Favata \cite{Favata2011}.

For Kerr BH with slow rotation ($a \ll M$) we have obtained the corrections up to quadratic terms \cite{Jefremov2015}:
\begin{equation}
\begin{aligned}
J_{\mathrm{\, ISCO}}&= \mp 2\sqrt{3}M -\frac{2 \sqrt{2}}{3}a +\frac{\sqrt{2}}{3}s \pm\frac{11}{36 \sqrt{3}}\frac{a}{M}s \pm\\&\pm \frac{4 \sqrt{3} M}{27}\left( \frac{a}{M} \right)^2 \pm \frac{1}{4 M \sqrt{3}}s^2,\\
E_{\mathrm{\, ISCO}}&= \frac{2 \sqrt{2}}{3} \pm\frac{1}{18 \sqrt{3}}\frac{a}{M} \pm \frac{1}{36 \sqrt{3}}\frac{s}{M} -\frac{\sqrt{2}}{81}\frac{a}{M}\frac{s}{M} -\\&-\frac{5}{162 \sqrt{2}}\left( \frac{a}{M} \right)^2 -\frac{5}{432 \sqrt{2}M^2}s^2 ,\\
r_{\mathrm{\, ISCO}}&= 6 M \pm 4 \sqrt{\frac{2}{3}}a \pm 2 \sqrt{\frac{2}{3}}s + \frac{2}{9} \frac{a}{M}s -\\&-\frac{7M}{18}\left( \frac{a}{M} \right)^2 -\frac{29}{72M}s^2 .\\
\Omega_{\mathrm{\, ISCO}} &= \mp \frac{1}{6 \sqrt{6}M} +\frac{11}{216 M}\frac{a}{M} +\frac{1}{48 M^2}s \, \mp\\ 
&\mp \left( \frac{1}{18 \sqrt{6}M}\frac{as}{M^2} + \frac{59}{648 \sqrt{6}M} \frac{a^2}{M^2} + \right. \\
&\left. + \frac{97}{3456 \sqrt{6}M}\frac{s^2}{M^2} \right).
\end{aligned}
\label{Kerr-slow1}
\end{equation}

For extreme Kerr BH ($a=M$) for counterrotation we have obtained \cite{Jefremov2015}: 
\begin{equation}
\begin{aligned}
J_{\mathrm{\, ISCO}}&= -\frac{22 \sqrt{3}}{9}M +\frac{82\sqrt{3}}{243}s,\\
E_{\mathrm{\, ISCO}}&=\frac{5 \sqrt{3}}{9} +\frac{\sqrt{3}}{243}\frac{s}{M},\\
r_{\mathrm{\, ISCO}}&= 9M +\frac{16}{9}s,\\
\Omega_{\mathrm{\, ISCO}} &= -\frac{1}{26M} +\frac{3s}{338 M^2}.\\
\end{aligned}
\label{Kerr-counter}
\end{equation}

For the case of corotation we have considered nearly extreme Kerr BH with $a=(1-\delta)M$ with $\delta \ll 1$, and have obtained:
\begin{equation}
\begin{aligned}
J_{\mathrm{\, ISCO}}&=\left( \frac{2}{\sqrt{3}} +\frac{2 \times 2^{2/3} \delta^{1/3}}{\sqrt{3}} \right)M +\\
&+\left( -\frac{2}{\sqrt{3}} +\frac{4 \times 2^{2/3}\delta^{1/3}}{\sqrt{3}} \right)s,\\
E_{\mathrm{\, ISCO}}&=\left( \frac{1}{\sqrt{3}} +\frac{2^{2/3} \delta^{1/3}}{\sqrt{3}} \right) +\\
&+\left( -\frac{1}{\sqrt{3}} +\frac{2 \times 2^{2/3}\delta^{1/3}}{\sqrt{3}} \right) \frac{s}{M},\\
r_{\mathrm{\, ISCO}}&= \left( 1 +2^{2/3}\delta^{1/3} \right)M -2 \times 2^{2/3} \delta^{1/3}s,\\
\Omega_{\mathrm{\, ISCO}} &= \frac{1}{2M}  - \frac{3 \times 2^{2/3} \delta^{1/3}}{8M} + \frac{9 \times 2^{2/3} \delta^{1/3}}{16M^2} s .\\
\end{aligned}
\label{Kerr-co}
\end{equation}
We see that in the case of $a=M$ ($\delta=0$) the corrections, linear in spin, are absent in formulae for ISCO radius and frequency. This was also demonstrated in \cite{Abramowicz1979}. In the work \cite{TanakaMino1996} on basis of the numerical calculation, it was noticed that in the extreme Kerr background for the parallel case the magnitude of test-body's spin does not influence the radius of the last stable orbit and it always remains equal to $M$. We have succeeded in proving this analytically. We have obtained the exact (in spin) values of ISCO parameters for nearly extreme Kerr BH in case of corotation \cite{Jefremov2015}:
\begin{equation}
\begin{aligned}
&J_{\mathrm{\, ISCO}}= 2 M E_{\mathrm{\, ISCO}} \, ,\\
&E_{\mathrm{\, ISCO}}= \frac{M^2-s^2}{M^2\sqrt{3 +6s/M}} +\\
&+ \frac{(M^2 -s^2)^{1/3}(2M +s)^{2/3} Z(M,s)^{2/3}}{\sqrt{3}M^{5/2}(M +2s)^{3/2}}  \, \delta^{1/3},\\
&r_{\mathrm{\, ISCO}}=M +\frac{M(M^2 -s^2)^{1/3}(2M +s)^{2/3}}{Z(M,s)^{1/3}}  \,  \delta^{1/3}, \\
&\Omega_{\mathrm{\, ISCO}} = \frac{1}{2M} -\\
&-\frac{3(M -s)^{1/3}(M +2s)}{4(2M +s)^{1/3}(M +s)^{2/3} Z(M,s)^{1/3}}\,  \delta^{1/3} ,\\
&Z(M,s) \equiv M^4 +7M^3s +9M^2 s^2 +11M s^3 -s^4 .\\
\end{aligned}
\label{Kerr-delta}
\end{equation}
From this solution we see that for $\delta=0$ ($a=M$) the radius and the angular frequency are independent of the particle's spin $s$ while the values of energy and total angular momentum depend on it.

It can be easily seen from the exact solution (\ref{Kerr-delta}) that for extreme Kerr BH solution for energy and angular momentum diverges with $s \rightarrow -M/2$. It shows that an approximation of test body does not work with such large values of $s$. Of course, limits of test body application depend on $a$, but we emphasize that all results beyond the approximation $s \ll M$ should be considered with big care, see \cite{SaijoMaeda1998}, \cite{Jefremov2015}.

For a spinless particle the conserved quantity is the orbital angular momentum $L_z$, whereas in the case of a spinning particle the conserved quantity is the total angular momentum $J_z$, which includes spin terms \cite{SaijoMaeda1998}. In this case 'orbital angular momentum' at infinity $L_z$ can also be introduced as $L_z=J_z - s$, see \cite{SaijoMaeda1998}.

In the paper \cite{Jefremov2015} we present formulae for a small-spin linear corrections for $E$, $J$ and $\Omega$ at circular orbit with a given radius $r$. It can be seen that for $r \rightarrow \infty$ the total angular momentum equals to $J=J_0 + s$, where $J_0$ is total angular momentum for non-spinning particle. In spinless case it consists of orbital angular momentum part only, therefore $J_0=L$. This justifies the introduction of orbital momentum at infinity just as difference between $J$ and $s$.

For circular orbits at finite radius (in particular, ISCO) orbital angular momentum cannot be defined by such simple way \cite{SaijoMaeda1998}. But using test body approximation $s \ll M$ allows us to use term 'corotation' and 'counterrotation' as terms for orbital motion.

If we use tentatively the orbital angular momentum in the form $L_z=J_z - s$ for ISCO, we will get for the Schwarzschild case:
\begin{equation}
L_{\mathrm{\, ISCO}} = 2 \sqrt{3} M - \left(1- \frac{\sqrt{2}}{3} \right) s_J  .
\label{L-orb}
\end{equation}
We see from (\ref{L-orb}) and (\ref{Schw}) that increasing of positive $s_J$ leads to increasing of $r_{\mathrm{\, ISCO}}$ and decreasing of orbital angular momentum $L_{\mathrm{\, ISCO}}$. At the same time the total angular momentum becomes bigger but it happens only due to increasing of its spin part.

\section{Binding energy in the innermost stable circular orbit}

Let us denote efficiency $\varepsilon$ as the fraction of the rest mass
energy that can be released in making the transition from rest at infinity to
the innermost stable circular orbit \cite{Hobson}, in our units it is given by
\begin{equation}
\varepsilon = 1 - E_{\mathrm{\, ISCO}} = E_{\mathrm{\, bind}}.
\end{equation}
Note that in our notations $E$ is energy per unit particle rest mass. In other words the efficiency is the binding energy $E_{\mathrm{\, bind}}$ at ISCO per unit mass. The efficiency shows how much energy can be released by radiation during the accretion process. For the Schwarzschild black hole and non-spinning test particle the efficiency equals to 0.057, and it reaches maximum for the extreme Kerr black hole -- 0.42 \cite{Hobson}, \cite{MTW}.

\begin{figure}
\centerline{\hbox{\includegraphics[width=0.45\textwidth]{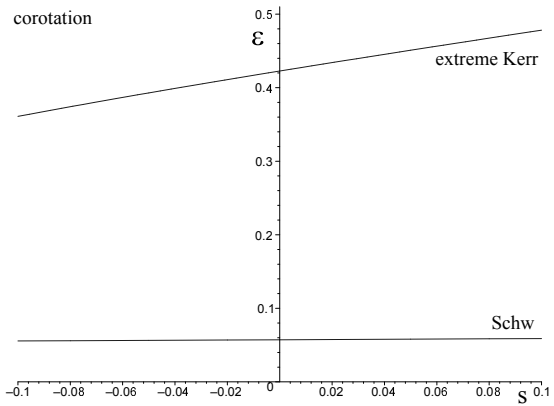}}} \caption{The efficiency (binding energy) of spinning test particle at ISCO. For case of extreme black hole the binding energy can be smaller or larger than 0.42 depending on spin orientation.} \label{fig-binding-energy}
\end{figure}

In the case of spinning test particles we can easily calculate the efficiency with using of expressions for $E_{\mathrm{\, ISCO}}$ presented in (\ref{Schw}), (\ref{Kerr-delta}), see Fig. \ref{fig-binding-energy}.

For the Schwarzschild black hole (see (\ref{Schw})) the efficiency is:
\begin{equation}
\varepsilon = 1 - \frac{2 \sqrt{2}}{3} + \frac{1}{36 \sqrt{3}}\frac{s_J}{M} .
\end{equation}
For extreme Kerr black hole in case of orbital corotation (see (\ref{Kerr-co})), the efficiency is:
\begin{equation}
\varepsilon = 1 - \frac{1}{\sqrt{3}} +  \frac{1}{\sqrt{3}}  \frac{s}{M} .
\end{equation}
It means that in case when particle spin is parallel to the total angular momentum of particle and the black hole spins, the efficiency can be larger than 42 \%. Note that the ISCO radius and angular frequency do not depend on spin in the case of extreme Kerr BH.

\section*{Acknowledgments}

The work of GSBK and OYuT was partially supported by the Russian Foundation for Basic Research Grant No. 14-02-00728 and the Russian Federation President Grant for Support of Leading Scientific Schools, Grant No.  NSh-261.2014.2. GSBK acknowledges partial support by by the Russian Foundation for Basic Research Grant No. OFI-M 14-29-06045.

\end{document}